\documentclass[aps,11pt,nofootinbib,preprintnumbers,showpacs]{revtex4}
\usepackage{amssymb, amsfonts, amsmath, bm}
\usepackage
{graphicx,color}

\begin{document}

\title{{\Large
Gravitational two solitons in Levi-Civit\`a spacetime}}

\hfill{\small RUP-15-25}

\author{Takahisa Igata$^{1}$}
\email{igata@rikkyo.ac.jp}

\author{Shinya Tomizawa$^2$}
\email{tomizawasny@stf.teu.ac.jp}

\affiliation{
${}^1$ Department of Physics, Rikkyo University, Toshima, Tokyo 175-8501, Japan\\
${}^2$ Department of Liberal Arts, Tokyo University of Technology, Otaku, Tokyo 144-8535, Japan}


\begin{abstract}
Applying the Pomeransky inverse scattering method to 
the four-dimensional vacuum Einstein equations 
and using the Levi-Civit\`a solution as a seed, 
we construct a two-soliton solution with cylindrical symmetry. 
In our previous work, we constructed the one-soliton solution with a real pole 
and showed that the singularities that the Levi-Civit\`a background has 
on an axis can be removed 
by the choice of certain special parameters, 
but it still has unavoidable null singularities, 
as usual one-solitons do. 
In this work, we show that for the two-soliton solutions, 
any singularities can be removed by suitable parameter-setting 
and such solutions describe the propagation of gravitational wave packets. 
Moreover, in terms of the two-soliton solutions, 
we mention a time shift phenomenon, 
the coalescence and the split of solitons 
as the nonlinear effect of gravitational waves. 
\end{abstract}

\maketitle

\section{Introduction}
\label{sec:1}
Gravitational solitons in general relativity describe gravitational solitonic waves 
propagating in a certain background spacetime. 
In particular, the use of inverse scattering method leads to
a lot of discoveries of various physically interesting 
exact solutions~\cite{book exact solution,Belinski:2001ph}. 
It should be especially mentioned that, 
in addition to exact solutions describing nonlinear
gravitational waves on various physical backgrounds, 
this method can generate black hole solutions 
in an axisymmetric and stationary case, 
and multiple-soliton solutions with stationary and axial symmetry, 
such as the generalized soliton solutions of the Weyl class, Kerr-NUT solutions, 
double Kerr metric, and the rotating Weyl $C$-metric \textit{etc}.~\cite{Belinski:2001ph, Letelier:1985, Carot:1989it, Chaudhuri:1997}.
This method can be generalized to 
the higher-dimensional Einstein equations~\cite{Iguchi:2011qi,review ER}, 
but in general such a simple generalization to higher dimensions 
tends to lead to singular solutions. 
However, Pomeransky~\cite{Pomeransky:2005sj} modified the original 
inverse scattering method~\cite{Belinsky:1979mh} 
so that it can generate regular solutions even 
in higher dimensions. 
In fact, with regard to vacuum solutions in five dimensions, 
all currently known asymptotically flat black hole solutions 
can be discovered or rederived 
by this method~\cite{Iguchi:2011qi,review ER}.

\medskip
In a cylindrically symmetric case, 
a diagonal form of a metric makes the vacuum Einstein equations 
 extremely simple structure of a linear wave equation in a flat background. 
For instance, 
the Einstein-Rosen metric can be interpreted as 
superposition of cylindrical gravitational waves 
with the $+$ mode only~\cite{Einstein-Rosen, book exact solution}. 
However, the existence of off-diagonal nonzero components 
of a metric drastically changes the structure of the Einstein equations and 
it yields the $\times$ mode together with nonlinearity. 
Piran \textit{et al}.~\cite{Piran} numerically studied the nonlinear interaction 
of cylindrical gravitational waves with both polarization modes 
such as the gravitational Faraday effect. 
Tomimatsu~\cite{Tomimatsu:1989vw} studied the gravitational Faraday rotation 
for the cylindrical gravitational solitons generated 
by the inverse scattering technique~\cite{Belinsky:1979mh}. 
As one of new attempts to understand strong gravitational effects, 
one of the present authors
 has recently constructed 
the cylindrically symmetric soliton solutions 
from a Minkowski seed by the Pomeransky 
inverse scattering method, 
and has clarified the behavior of the soliton solution including 
the effect similar to the gravitational Faraday rotation~\cite{Tomizawa:2013soa,Tomizawa:2015zva}.

\medskip
In our previous paper, using the Pomeransky inverse scattering method 
and regarding the Levi-Civit\`a metric as a seed, 
we constructed the one-soliton 
solution that does not admit staticity but cylindrical symmetry~\cite{Igata:2015oea}. 
Although Levi-Civit\`a spacetime has singularities on the axis except for Minkowski spacetime, 
we showed that for the one-soliton solution, 
such singularities disappear by a certain choice of parameters. 
However, this solution has singularities on the light cone for any parameters, 
which is an unavoidable common property for all
one-soliton solutions with a real pole and it is well known that
such a problem 
can be resolved by considering two-soliton solutions 
with two complex-conjugate poles. 
Therefore, in this paper, 
we will construct a two-soliton solution with two complex 
conjugate poles by the Pomeransky method and the Levi-Civit\`a seed. 
Furthermore, 
we will show that for certain parameters, 
the spacetime is entirely free from any singularities 
even on the axis as well as on the light cone.

\medskip
In the following section, 
we will present a two-soliton solution with complex conjugate poles 
in Levi-Civit\`a spacetime. 
In Sec.~\ref{sec:3} 
we will analyze the obtained two-soliton solution 
by computing the amplitudes and polarization angles for ingoing and outgoing waves
 and will see the difference from the one-soliton solution. 
In Sec.~\ref{sec:4}, we will give the summary and discussion on our results.

\section{Two-soliton solution}
\label{sec:2}
The most general cylindrically symmetric solution to 
the four-dimensional vacuum Einstein equations
can be described in the Kompaneets-Jordan-Ehlers form~\cite{Komaneets-Jordan-Ehlers} 
\begin{align}
ds^2
=e^{2\psi}\left(dz+\omega\,d\phi\right)^2
+\rho^2\,e^{-2\psi}\,d\phi^2
+e^{2\left(\gamma-\psi\right)}\left(d\rho^2-dt^2\right),
\label{eq:metric}
\end{align}
where the functions $\psi$, $\omega$, and $\gamma$ 
depend on the time coordinate $t$ and radial coordinate $\rho$ only. 
Let us define a $2\times 2$ metric $g$ and a metric function $f$ by
\begin{align}
&g=\left(
\begin{array}{cc}
e^{2\psi}& \omega \,e^{2\psi} \\
\omega\,e^{2\psi}&\rho^2\,e^{-2\psi}+\omega^2e^{2\psi}
\end{array}
\right),
\\
&f=e^{2\left(\gamma-\psi\right)}, 
\end{align}
respectively.

As shown by Belinsky and Sakharov~\cite{Belinsky:1979mh}, 
the vacuum Einstein equation with cylindrical symmetry (in general, 
with two commuting Killing vectors) is completely integrable, 
and it admits a pair of linear equations, 
which is called Lax pair
\begin{align}
D_\rho\Psi=\frac{\rho\, U+\lambda\, V}{\rho^2-\lambda^2}\Psi,\quad D_t\Psi=\frac{\rho\, V+\lambda \,U}{\rho^2-\lambda^2}\Psi,
\end{align}
where $\lambda$ is a spectral parameter (complex parameter), 
$D_\rho$ and $D_t$ are the commuting differential operators defined by
\begin{align}
D_\rho:=\partial_\rho+\frac{2\lambda\rho}{\rho^2-\lambda^2}\partial_\lambda,\quad D_t:=\partial_t+\frac{2\lambda^2}{\rho^2-\lambda^2}\partial_\lambda, 
\end{align}
and  $U$ and $V$ are the $2\times 2$ matrices defined by $U=\rho\, g_{,\rho}\,g^{-1}$ and $V=\rho \,g_{,t}\,g^{-1}$, respectively, and  the generating matrix $\Psi(\lambda,\rho,t)$ is a $2\times2$ matrix such that $\Psi(\lambda=0,\rho,t)=g$. Therefore, one can obtain the $2\times2$ metric $g$ from the generating matrix $\Psi(\lambda,\rho,t)$ (more precisely, by normalizing the metric such that $\mathrm{det}\ g=\rho^2$).

\medskip
To generate new cylindrically symmetric solutions, we start from one solution of the Lax pair such that $\Psi_{0}(\lambda=0,\rho,t)=G_{0}$ ($G_0$ : a seed metric) and then must dress it as
\begin{align}
\Psi=\chi\Psi_0,
\end{align}
where in particular, for a two-soliton solution, the dressing matrix $\chi$ is given by 
\begin{eqnarray}
\chi&=&1+\sum_{k=1,2}\frac{R_k}{\lambda-\mu_k},
\end{eqnarray}
with
\begin{align}
&(R_k)_{ab}=m_a^{(k)}\sum_{l=1,2}\frac{(\Gamma^{-1})_{kl}m_c^{(l)}G_{0cb}}{\mu_l},\\
&\Gamma_{kl}=\frac{m^{(k)}_a G_{0ab}\,m^{(l)}_b}{-\rho^2+\mu_k\mu_l},\\
&m^{(k)}_a=m^{(k)}_{0b}\left[\Psi^{-1}_{0}(\mu_k,\rho,t)\right]_{ba},
\end{align}
where, 
following the notation in \cite{Belinski:2001ph}, we use the successive Latin subscripts to denote the summation of the $z$ and $\phi$ components, and 
$\mu_i$ $(i=1, 2)$ are defined by
\begin{align}
&\mu_i=\sqrt{(t-a_i)^2-\rho^2}-\left(t-a_i\right).
\end{align}
The constant vectors $m^{(k)}_{0}=(m_{0z}^{(k)},m_{0\phi}^{(k)})\ (k=1,2)$ add the seed to new parameters (the obtained solution is invariant under the transformation $m_0^{(k)}\to k\,m_0^{(k)}$ ($k$: a nonzero constant)). 
After dressing, taking the limit of $\lambda\to0$, one can obtain the metric $g=\Psi(\lambda=0,\rho,t)$. 
In the Belinsky-Zakarov method, one must normalize the obtained metric. 
Furthermore, for reality of the metric, one must put $a_1=\bar a_2$.

In this paper, following the Pomeransky method~\cite{Pomeransky:2005sj}, in which 
one need not normalize the metric obtained from the dressed generating matrix.  
For the Levi-Civit\`a metric, which we choose as a seed, 
the $2\times2$ metric $g_0$ and the metric function $f_0$ are written as
\begin{align}
&g_0=\mathrm{diag}\left(
\rho^{1-d},\, \rho^{1+d}\right),
\\
&f_0=b^2 \rho^{(d^2-1)/2},
\end{align}
respectively, where the parameters $b$ and $d$ are independent, 
and are assumed to be positive without loss of generality. 
Let us remove trivial solitons at $t=a_1$ and $t=a_2$ with a vector $(1,0)$ (this corresponds to the inverse transformation $g_0'=\chi(\lambda=0,\rho,t)^{-1}g_0$ with $m_{0}^{(k)}=(1,0)\ (k=1,2)$  in the Belinski-Zakharov method), and then we have the metric
\begin{align}
g'_{0}
=\mathrm{diag}
\left(\rho^{-3-d}\mu_1^2\,\mu_2^2,\, \rho^{1+d}\right)
=
\mathrm{diag}\left(
\frac{\rho^{5-d}}{\tilde{\mu}_1^2\,\tilde{\mu}_2^2},\, \rho^{1+d}\right),
\end{align}
and dress the generating matrix $\Psi_0$ corresponding to the metric $g'_0$:
\begin{align}
\Psi_0(\lambda, \rho, t)
=\mathrm{diag}\left(\frac{\left(\rho^2+2\,t\, \lambda+\lambda^2\right)^{\left(5-d\right)/2}}
{\left(\tilde\mu_1-\lambda\right)^2\left(\tilde\mu_2-\lambda\right)^2},\,(\rho^2+2\,t\, \lambda+\lambda^2)^{\left(1+d\right)/2}\right),
\end{align}
where 
\begin{align}
\tilde{\mu}_i=\frac{\rho^2}{\mu_i}=-\sqrt{(t-a_i)^2-\rho^2}-\left(t-a_i\right).
\end{align}
Next, add back nontrivial solitons with BZ vectors 
$m^{(1)}_{0}=(1,a)$ and $m^{(2)}_0=(1,\bar a)$, 
and then we obtain a two-soliton solution as
\begin{align}
&g_{ab}
=g'_{0ab}
-\sum_{k,l=1}^2\frac{g'_{0ac}\,m^{(k)}_c\,(\Gamma^{-1})_{kl}\,m^{(l)}_d\,g'_{0db}}{\mu_k\mu_l},
\\
&f=f_0\,\frac{\mathrm{det}\,\Gamma_{kl}}{\mathrm{det}\,\Gamma_{0kl}},
\end{align}
where
$\Gamma_{0kl}$ is $\Gamma_{kl}$ evaluated at $a=0$.

\medskip
This is how we can obtain a two-soliton solution, 
whose metric can be written 
in the Kompaneets-Jordan-Ehlers form~\eqref{eq:metric}, 
where the functions $\psi$, $\omega$ and $\gamma$ are explicitly written as
\begin{align}
&
e^{2\psi}
=\rho^{1-d}\,|\,w\,|^4\left(1-\frac{\mathcal{A}}{\mathcal{B}}\right),
\\
&
\omega
=-\frac{\left(|\,w\,|^2-1\right)^2}{\rho^{2-d}}\frac{\mathcal{C}}{\mathcal{B}-\mathcal{A}},\\
&
e^{2\gamma}
=b^2\rho^{\left(d-1\right)^2/2}\,\frac{|\,w\,|^{4}\left(\mathcal{B}-\mathcal{A}\right)}{\left(w-\bar{w}\right)^2|\,X\,|^3},
\end{align}
with\footnote{
The function $\omega$ can be rewritten in terms of $X$ and $Y$ as
\begin{align*}
\omega
=-\frac{2\,\rho^{d}}{\mathcal{B}-\mathcal{A}}\,
\Re\,\left[\,
\frac{\bar{c}\,\bar{Y}}{|\,w\,|^2}\left(
\frac{\bar{X}}{\bar{w}^{2}}\,
\frac{X^2+c^2\,Y^2}{w^2-1}
-
\frac{X}{|\,w\,|^2}\,\frac{|\,X\,|^2+|\,c\,|^2\,|\,Y\,|^2}{|\,w\,|^2-1}
\right)
\,\right].
\end{align*}}
\begin{align}
&\mathcal{A}
=2\,\Re\left[\,
\frac{\bar{X}^2}{\bar{w}^{2}}\,
\frac{X^2+c^2\,Y^2}{w^2-1}
\,\right]
-2\,\frac{|\,X\,|^2}{|\,w\,|^2}\,\frac{|\,X\,|^2+|\,c\,|^2\,|\,Y\,|^2}{|\,w\,|^2-1},
\\
&\mathcal{B}
=\frac{|\,X^2+c^2\,Y^2\,|^2}{|\,w^2-1\,|^2}
-\frac{\left(|\,X\,|^2+|\,c\,|^2\,|\,Y\,|^2\right)^2}{\left(|\,w\,|^2-1\right)^2},
\\
&
\mathcal{C}=2\,\Re\,\left[\frac{\bar{c}\left(\bar w^2-1\right)^2}{\bar{w}^d\left(w^2-1\right)}\left(X^2+c^2\,Y^2\right)\right]
-2\,\Re\,\left[\frac{\bar{c}\left(w^2-1\right)^2}{\bar{w}^{d-1}w\left(|\,w\,|^2-1\right)}\right]\left(|\,X\,|^2+|\,c\,|^2\,|\,Y\,|^2\right),
\end{align}
\begin{align}
&X=\left(w^2-1\right)^2\left(|\,w\,|^2-1\right)^2,
\\
&Y=\frac{|\,w\,|^2}{\rho^2}\,w^{2-d},
\\
&w=\frac{\mu_1}{\rho},
\\
&c=\left(2\,a_1\right)^{2-d}\,a.
\end{align}
Here, $\Re[\ ]$ denotes the real part of $[\ ]$. 
When $d=1$, we recover the metric in \cite{Tomizawa:2015zva}, 
in which case the seed is a Minkowski metric, 
and when $c=0$, 
this metric reduces to the Levi-Civit\`a metric.

\medskip
A shift of the time coordinate allows us to fix the parameters $a_i$ as a pure imaginary 
\begin{align}
a_1=\bar{a}_2=i\,q,
\end{align}
where $i$ denotes the imaginary unit 
and $q$ is assumed to be positive without 
loss of generality.

\section{Analysis for two-solitons}
\label{sec:3} 

\subsection{Asymptotic behaviors of gravitational waves}
In this subsection, we analyze the asymptotic behaviors of gravitational waves 
at spacetime boundaries 
for the obtained two-soliton solution 
by computing the asymptotic behaviors of the metric 
and ingoing and outgoing wave amplitudes, 
where we basically follow the definitions given 
in Refs.~\cite{Piran, Tomimatsu:1989vw} (see Appendix~\ref{sec:A}). 
First, let us consider the symmetric axis $\rho=0$. 
In order to focus on the physical propagation process 
of cylindrically symmetric gravitational soliton waves, 
we investigate the condition for the parameters $c$ and $d$ 
such that there do not exist any gravitational sources on the axis, 
i.e., the \textit{C}-energy density could become finite on the axis.

\medskip
Note that the \textit{C}-energy density is proportional to $\gamma_{,\rho}$. 
Near $\rho\simeq 0$, the metric function $e^{2\gamma}$ behaves asymptotically as 
{\small 
\begin{align}
e^{2\gamma}\simeq 
b^2\bigg[\,
\rho^{\left(d-1\right)^2/2}
-\frac{4^{d-4}}{q^2}
\frac{\Im\,\left[\,
c\,(-t-i\,q)^{3-d}\,\right]^2}{\left(t^2+q^2\right)^{2\left(3-d\right)}}
\rho^{\left(d-3\right)^2/2}
+16^{d-5}|\,c\,|^4\left(t^2+q^2\right)^{2\left(d-5\right)}\rho^{\left(d-5\right)^2/2}
\,\bigg],
\end{align}}
where $\Im\,[\ ]$ denotes the imaginary part of $[\ ]$.

\medskip
For $d\not =1,\ 3,\ 5$, the \textit{C}-energy density diverges on $\rho= 0$ as
\begin{align}
\gamma_{,\rho}=O(\rho^{-1}), 
\end{align}
but for $d=1,\  3, \ 5$, it vanishes as 
\begin{alignat}{2}
&\gamma_{,\rho}\simeq 
\dfrac{\left[\,
c_\textrm{I}\left(t^2-q^2\right)+2\,q\,c_\textrm{R}t\,\right]^2
}{64\,q^2\left(t^2+q^2\right)^4}
\,\rho, 
~~~&\textrm{for} ~~ d=1,
\\
&\gamma_{,\rho}\simeq 
\left[\,
\dfrac{4\,q^2}{c_\textrm{I}^2}\left(
1+\dfrac{|\,c\,|^4}{256\left(t^2+q^2\right)^{4}}\right)
+
\frac{q^2\,|\,c\,|^2+2\,c_\textrm{I}\left[\,
c_\textrm{I}\left(t^2-q^2\right)+2\,q\,c_\textrm{R}\,t
\,\right]}{2\,c_\textrm{I}^2\left(t^2+q^2\right)^2}\,\right]
\rho, 
~~~&\textrm{for} ~~ d=3,
\\
&\gamma_{,\rho}\simeq
\left[\,
\dfrac{4\left[\,
c_\textrm{I}\left(t^2-q^2\right)-2\,q\,c_\textrm{R}\,t
\,\right]^2
}{q^2\,|\,c\,|^4}+\frac{4\,t^2}{\left(t^2+q^2\right)^2}\,\right]\,\rho,
~~~&\textrm{for} ~~ d=5,
\end{alignat}
where we assume $c_\textrm{I}\neq0$ 
 for $d=3$ because if not, $\gamma_{,\rho}$ diverges on $\rho= 0$. 
The case of $d=1$ (two-soliton solution obtained from Minkowski metric) 
have been previously analyzed by one of the authors~\cite{Tomizawa:2015zva}. 
This is why we focus on $d=3\, (c_\textrm{I}\neq0)$ and $d=5$ only in what follows.

\medskip
Near $\rho=0$, the metric behaves as
\begin{alignat}{3}
ds^2
\simeq \,
& \dfrac{4\,c_\textrm{I}^2\left(t^2+q^2\right)^{2}}{q^2|\,c\,|^4+16\,\left(t^2+q^2\right)^{2}\left(c_\textrm{I} t-q\,c_\textrm{R}\right)^2}\,d\tilde{z}^2
&&
\nonumber
\\
&+
\dfrac{q^2|\,c\,|^4+16\,\left(t^2+q^2\right)^{2}\left(c_\textrm{I} t-q\,c_\textrm{R}\right)^2}{4\,c_\textrm{I}^2\left(t^2+q^2\right)^{2}}
\,\left[\,
\rho^2\,d\tilde{\phi}^2
+\dfrac{b^2c_\textrm{I}^2}{4\,q^2}\left(d\rho^2-dt^2\right)
\,\right],
~~~&\textrm{for} ~~&d=3,
\\
ds^2
\simeq \,&\dfrac{1}{16\left(t^2+q^2\right)^2}\left(dz
-\frac{32\left(t^2+q^2\right)^2\left[\,
c_\textrm{I}\left(t^2-q^2\right)-2\,q\,c_\textrm{R} t
\,\right]}{q\,|\,c\,|^2}\,\rho^2\,d\phi
\right)^2
&&
\nonumber
\\
&+16\left(t^2+q^2\right)^2\left[\,
\rho^2\,d\phi^2+\,b^2\,|\,c\,|^4\left(d\rho^2-dt^2\right)
\,\right],
~~~& \textrm{for} ~~&d=5,
\end{alignat}
where we have introduced the new coordinates 
$\tilde{z}=z-(2\,q\,|\,c\,|^2/c_\textrm{I})\,\phi$ 
 and $\tilde{\phi}=\phi$ 
for $d=3$. 
A deficit angle $\Delta$ on the axis is defined as
\begin{align}
\Delta \equiv 2\pi-\lim_{\rho\to 0}\dfrac{\int_0^{\Delta\phi}\sqrt{g_{\phi\phi}}\,d\phi}{\int_0^\rho \sqrt{g_{\rho\rho}}\,d\rho},
\end{align}
where 
$\phi$ is replaced by $\tilde{\phi}$ for $d=3$, and
$\Delta \phi$ is the periodicity of the angular coordinate. 
Then we have 
\begin{alignat}{2}
&\Delta=2\pi\left(
1-\frac{q\,\Delta \phi}{\pi b \,|\,c_\textrm{I}\,|}
\right),
~~~&\textrm{for} ~~d=3,
\\
&
\Delta=2\pi\left(1-
\frac{\Delta \phi}{2\pi b\,|\,c\,|^2}
\right),
~~~&\textrm{for} ~~d=5.
\end{alignat}
By choosing a suitable value of $\Delta \phi$ to be 
$\Delta=0$, 
we get rid of conical singularities on the axis.

\medskip
Near $\rho=0$, 
the ingoing and outgoing total amplitudes, $A$ and $B$, behave as {\small
\begin{alignat}{2}
&A\simeq B\simeq 
\frac{\sqrt{q^2|\,c\,|^4+32\,(t^2+q^2)^2\big[\,
q^2\,(c_\textrm{R}^2-c_\textrm{I}^2)+4\,q\,c_\textrm{I} c_\textrm{R}\,t+2\,c_\textrm{I}^2\,t^2
+8\,q^2\,(t^2+q^2)^2
\,\big]}
}{4\,|c_\mathrm{I}|\left(t^2+q^2\right)^2},
~~~&\textrm{for} ~~d=3,
\\
&A\simeq B\simeq \frac{4\sqrt{q^2\,|\,c\,|^4\,t^2
+(t^2+q^2)^2\,[\,c_\mathrm{I}\left(t^2-q^2\right)-2\,q\,c_\mathrm{R}\,t\,]^2
}}{q\,|\,c\,|^2\left(t^2+q^2\right)},
~~~&\textrm{for} ~~d=5,
\end{alignat}}
respectively. 
The polarization angles, $\theta_A$ and $\theta_B$, take the following asymptotic forms
{\footnotesize
\begin{alignat}{2}
\tan 2\theta_A
&\simeq
-\tan 2\theta_B
\nonumber
\\
&\simeq\frac{q^3|\,c\,|^6
+16\,q\,(t^2+q^2)^2\big(\,
16\,(q\,c_\textrm{R}-c_\textrm{I}\,t)^2(t^2+q^2)^2
-|\,c\,|^2\,[\,c_\textrm{I}^2(q^2+5\,t^2)-2\,q\,c_\textrm{R}(q\,c_\textrm{R}+t\,c_\textrm{I})\,]
\,\big)}{16\,c_\mathrm{I}\left(t^2+q^2\right)\left[\,
q^2\,|\,c\,|^4\,t+8\,c_\mathrm{I}\left(q\,c_\mathrm{R}-c_\mathrm{I}\,t\right)\left(t^2+q^2\right)^3
\,\right]},
~~~&\textrm{for} ~~d=3,
\\
\tan 2\theta_A
&\simeq -\tan 2\theta_B\simeq 
\frac{\left(t^2+q^2\right)\left[c_\mathrm{I}\left(t^2-q^2\right)-2\,q\,c_\mathrm{R}\,t\right]}{q\,|\,c\,|^2\,t},
~~~&\textrm{for} ~~d=5.
\end{alignat}}

\medskip
At late time $t\to\infty$ or at early time $t\to-\infty$, the metric behaves as
{\small
\begin{alignat}{2}
ds^2\simeq
&
\frac{c_\mathrm{I}^2+4\,q^2\rho^2}{4\,c_\mathrm{I}^2\,t^2}\left(
dz+\frac{8\,q\,c_\mathrm{I}\,\rho^2\,t^2}{c_\mathrm{I}^2+4\,q^2\,\rho^2}\,d\phi
\right)^2
+
\frac{4\,c_\mathrm{I}^2\,t^2}{c_\mathrm{I}^2+4\,q^2\,\rho^2}\,\rho^2\,d\phi^2
+b^2\,\frac{c_\mathrm{I}^2\,t^2}{q^2}\,\left(d\rho^2-dt^2\right),
~~~& \mathrm{for} ~~ d=3,
\\
ds^2\simeq
&
\frac{1}{4\,\rho^2\,t^2}\left(
dz-\frac{8\,q\,|\,c\,|^2\,t^2}{c_\mathrm{I}}\,d\phi
\right)^2
+4\,\rho^4\,t^2\,d\phi^2
+b^2\frac{16\,c_\mathrm{I}^2\,\rho^4\,t^6}{q^2}\left(d\rho^2-dt^2\right),
~~~& \mathrm{for} ~~ d=5.
\end{alignat}}
The total amplitudes and polarization angles behave as
\begin{alignat}{2}
&A\simeq B \simeq
 \frac{4\,q}{\sqrt{c_\mathrm{I}^2+4\,q^2\rho^2}},
 ~~~& \mathrm{for} ~~ d=3,
\\
&A\simeq B\simeq 
\frac{2}{\rho}, 
~~~& \mathrm{for} ~~ d=5,
\label{eq:amplitude_latetime}
\end{alignat}
and
\begin{alignat}{2}
&\tan 2\theta_A\simeq \tan 2\theta_B\simeq \frac{c_\mathrm{I}}{2\,q\,\rho},
~~~& \mathrm{for} ~~ d=3,
\\
&\tan 2\theta_A\simeq -\tan 2\theta_B\simeq 
\frac{2\,q\,|\,c\,|^2}{c_\mathrm{I}\,\rho^2\,t},
~~~& \mathrm{for} ~~ d=5.
\end{alignat}

\medskip
At spacelike infinity $\rho\to\infty$, the metric behaves as
{\small
\begin{alignat}{2}
ds^2\simeq
&
\frac{1}{\rho^2}
\left(
dz-\frac{64\,q^3\,c_\mathrm{I}\,\rho^2}{|\,c\,|^2+256\,q^4}\,d\phi
\right)^2
+\rho^4
\,d\phi^2
+\frac{b^2\left(|\,c\,|^2+256\,q^4\right)^2}{\left(256\,q^4\right)^2}\,\rho^4
\left(d\rho^2-dt^2\right),
~~~&\mathrm{for} ~~ d=3,
\\
ds^2\simeq
&
\frac{1}{\rho^4}
\left(
dz+\frac{64\,q^3\,c_\mathrm{I}\,\rho^4}{|\,c\,|^2+256\,q^4}\,d\phi
\right)^2
+\rho^6
\,d\phi^2
+\frac{b^2\left(|\,c\,|^2+256\,q^4\right)^2}{\left(256\,q^4\right)^2}\,\rho^{12}
\left(d\rho^2-dt^2\right),
~~~&\mathrm{for} ~~ d=5.
\end{alignat}
}
The total amplitudes and polarization angles behave as
\begin{alignat}{2}
&A\simeq B\simeq \frac{2}{\rho},
~~~& \mathrm{for} ~~ d=3,
\\
&A\simeq B\simeq 
\frac{4}{\rho},
~~~& \mathrm{for} ~~ d=5,
\end{alignat}
and
\begin{alignat}{2}
&\tan 2\theta_A\simeq 
\frac{32\,q^3\left(2\,c_\textrm{I}+3\,c_\textrm{R}\right)}{\left(|\,c\,|^2+256\,q^4\right)\rho}
,\quad 
\tan 2\theta_B
\simeq 
\frac{32\,q^3\left(2\,c_\textrm{I}-3\,c_\textrm{R}\right)}{\left(|\,c\,|^2+256\,q^4\right)\rho},
~~~
& \mathrm{for} ~~ d=3,
\\
&\tan 2\theta_A\simeq -\frac{16\,q^3\left(4\,c_\textrm{I}+5\,c_\textrm{R}\right)}{\left(|\,c\,|^2+256\,q^4\right)\rho},
\quad
\tan 2\theta_B
\simeq 
-\frac{16\,q^3\left(4\,c_\mathrm{I}-5\,c_\mathrm{R}\right)}{\left(|\,c\,|^2+256\,q^4\right)\rho},
~~~
& \mathrm{for} ~~ d=5.
\end{alignat}

\medskip
At past null infinity $u\to-\infty$, the metric behaves as
\begin{alignat}{2}
&ds^2\simeq \frac{1}{u^2}\left(dz+\omega_\textrm{p}\,\left(-u\right)^{5/2}\,d\phi\right)^2
+u^4\,d\phi^2+b^2f_\textrm{p}\,u^{4}\left(d\rho^2-dt^2\right),
~~~& \textrm{for} ~~d=3,
\\
&ds^2\simeq \frac{1}{u^4}\left(dz+\omega_\textrm{p}\,\left(-u\right)^{9/2}\,d\phi\right)^2
+u^6\,d\phi^2+b^2f_\textrm{p}\,u^{12}\left(d\rho^2-dt^2\right), 
~~~& \textrm{for} ~~d=5,
\end{alignat}
where $\omega_\textrm{p}$ and $f_\textrm{p}$ are defined by
\begin{align}
\omega_\textrm{p}
=
&\frac{64\,q^3\,p_v}{\mathcal{D}_\textrm{p}}\,\bigg[\,
256\,q^{11}\,c_\textrm{I}-256\,q^{10}\,c_\textrm{R}\,p_v^2
+768\,c_\textrm{I}\,q^9\,p_v^4
-768\,q^8\,c_\textrm{R}\,p_v^6
-q^3\,c_\textrm{I}\left(|\,c\,|^2-768\,q^4\right)p_v^8
\cr&
+3\,q^2\,c_\textrm{R}\left(|\,c\,|^2-256\,q^4\right)p_v^{10}
+q\,c_\textrm{I}\left(3\,|\,c\,|^2+256\,q^4\right)p_v^{12}
-c_\textrm{R}\left(|\,c\,|^2+256\,q^4\right)p_v^{14}
\,\bigg],
\\
f_\textrm{p}=&\frac{\mathcal{D}_\textrm{p}}{\big[\,256\,q^4\left(q^2+p_v^4\right)^2\,\big]^2}.
\end{align}

\medskip
The asymptotic forms of the amplitudes and the polarization angles are 
\begin{align}
&A \simeq \frac{8\,q\,p_v^3}{p_v^4+q^2}\,\sqrt{-\frac{\mathcal{N}_{\rm p}}{\mathcal{D}_{\rm p}u}}, \quad
B\simeq -\frac{2}{u},
 ~~~ \mathrm{for} ~~ d=3,
\label{eq:Apast-d3}
 \\
&A \simeq \frac{8\,q\,p_v^3}{p_v^4+q^2}\,\sqrt{-\frac{\mathcal{N}_{\rm p}}{\mathcal{D}_{\rm p}u}}, \quad
B\simeq -\frac{4}{u},
 ~~~ \mathrm{for} ~~ d=5,
\label{eq:Apast-d5}
\end{align}
where
\begin{align}
\mathcal{N}_{\rm p}=&\,
256\,q^{10}\,c_\mathrm{I}^2-
1536\,q^{9}c_\mathrm{R}\,c_\mathrm{I}\,p_v^2-
768\,q^{8}\left(2\,c_\mathrm{I}^2-3\,c_\mathrm{R}^2\right)p_v^{4}+
5120\,q^7\,c_\mathrm{R}\,c_\mathrm{I}\,p_v^{6}
\cr&
+
768\,q^6\left(3\,c_\mathrm{I}^2-2\,c_\mathrm{R}^2\right)p_v^{8}-
1536\,q^5\,c_\mathrm{R}\,c_\mathrm{I}\,p_v^{10}+
\left(|\,c\,|^4+256\,q^4\,c_\mathrm{R}^2\right)p_v^{12},
\\
\mathcal{D}_{\rm p}=&\,
65536\,q^{16}+
1024\,q^{10}\left(c_\mathrm{I}^2+256\,q^4\right)p_v^4-
4096\,q^{9}c_\mathrm{R}\,c_\mathrm{I}\,p_v^6+
1536\,q^{8}\left(-c_\mathrm{I}^2+3\,c_\mathrm{R}^2+256\,q^4\right)p_v^8
\cr&
+
4096\,q^{7}c_\mathrm{R}\,c_\mathrm{I}\,p_v^{10}+
1024\,q^{6}\left(2\,c_\mathrm{I}^2+c_\mathrm{R}^2+256\,q^4\right)p_v^{12}+
\left(
|\,c\,|^2+256\,q^4\right)^2p_v^{16},
\end{align}
and
\begin{align}
p_v=\sqrt{2v+\sqrt{4v^2+q^2}}. 
\end{align}

\medskip
At future null infinity $v\to\infty$, the metric behaves as 
\begin{alignat}{2}
&ds^2\simeq
\frac{1}{v^2}\left(dz+\omega_\textrm{f}\,v^{5/2}\,d\phi\right)^2+v^4\,d\phi^2+b^2f_\textrm{f}\,v^4\left(d\rho^2-dt^2\right), 
~~~& \textrm{for} ~~d=3,
\\
&ds^2\simeq
\frac{1}{v^4}\left(dz+\omega_\textrm{f}\,v^{9/2}\,d\phi\right)^2+v^{6}\,d\phi^2+b^2 f_\textrm{f}\,v^{12}\left(d\rho^2-dt^2\right), ~~~& \textrm{for} ~~d=5,
\end{alignat}
where $\omega_\textrm{f}$ and $f_\textrm{f}$ are defined by
\begin{align}
\omega_\textrm{f}
=
&\,
\frac{64\,q\,p_u}{\mathcal{D}_\textrm{f}}\,\bigg[\,
q^3\,c_\textrm{R}\left(|\,c\,|^2+256\,q^4\right)+q^2\,c_\textrm{I}\left(3\,|\,c\,|^2+256\,q^4\right)p_u^2
-3\,q\,c_\textrm{R}\left(|\,c\,|^2-256\,q^4\right)p_u^4
\cr&
-c_\textrm{I}\left(|\,c\,|^2-768\,q^4\right)p_u^6
+768\,q^3\,c_\textrm{R}\,p_u^8
+768\,q^2\,c_\textrm{I}\,p_u^{10}
+256\,q\,c_\textrm{R}\,p_u^{12}
+256\,c_\textrm{I}\,p_u^{14}\,\bigg],
~~~
\\
f_\textrm{f}
=&
\frac{\mathcal{D}_\textrm{f}}{\big[\,256\,q\left(q^2+p_u^4\right)^2\big]^2}. 
\end{align}

\medskip
The asymptotic forms of the amplitudes and the polarization angles are 
\begin{align}
&A\simeq \frac{2}{v},
\quad
B\simeq \frac{8\,p_u^3}{p_u^4+q^2}\,\sqrt{\frac{\mathcal{N}_{\rm f}}{\mathcal{D}_{\rm f}\,v}},
 ~~~ \mathrm{for} ~~ d=3,
\label{eq:Bfuture-d3}
\\
&A\simeq \frac{4}{v},
\quad
B\simeq \frac{8\,p_u^3}{p_u^4+q^2}\,\sqrt{\frac{\mathcal{N}_{\rm f}}{\mathcal{D}_{\rm f}\,v}},
~~~ \mathrm{for} ~~ d=5,
\label{eq:Bfuture-d5}
\end{align}
where $\mathcal{N}_{\rm f}$ and $\mathcal{D}_{\rm f}$ are defined by
\begin{align}
\mathcal{N}_{\rm f}=
&\,
q^2\left(|\,c\,|^4+256\,q^4\,c_\mathrm{R}^2\right)
+1536\,q^5\,c_\mathrm{R}\,c_\mathrm{I}\,p_u^2
+768\,q^4\left(3\,c_\mathrm{I}^2-2\,c_\mathrm{R}^2\right)p_u^4
\cr&
-5120\,q^3\,c_\mathrm{R}\,c_\mathrm{I}\,p_u^6
-768\,q^2\left(2\,c_\mathrm{I}^2-3\,c_\mathrm{R}^2\right)p_u^8
+1536\,q\,c_\mathrm{R}\,c_\mathrm{I}\,p_u^{10}
+256\,c_\mathrm{I}^2\,p_u^{12},
\\
\mathcal{D}_{\rm f}=
&\,q^2\left(|\,c\,|^2+256\,q^4\right)^2
+1024\,q^4\left(c_\mathrm{I}^2+|\,c\,|^2+256\,q^4\right)p_u^4
-4096\,q^3\,c_\mathrm{R}\,c_\mathrm{I}\,p_u^6
\cr&
+1536\,q^2\left[\,2\left(c_\mathrm{R}^2-c_\mathrm{I}^2\right)+|\,c\,|^2+256\,q^4\,\right]p_u^8
+4096\,q\,c_\mathrm{R}\,c_\mathrm{I}\,p_u^{10}
\cr&
+1024\left(c_\mathrm{I}^2+256\,q^4\right)p_u^{12}
+65536\,q^2\,p_u^{16},
\end{align}
and 
\begin{align}
p_u=\sqrt{2\,u+\sqrt{4\,u^2+q^2}}.
\end{align}

\subsection{Wave propagation}

\begin{figure}[!t]
\centering
\begin{tabular}{lll}
$n=1$
&
$n=3$
&
$n=4$
\\[-1mm]
\includegraphics[width=5cm,clip]{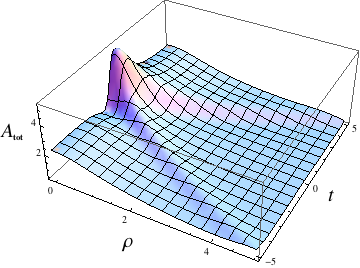}~~
&
\includegraphics[width=5cm,clip]{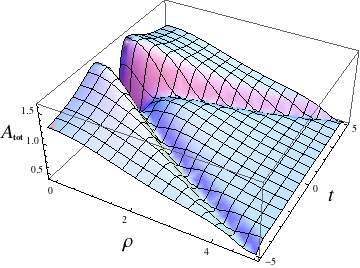}~~
&
\includegraphics[width=5cm,clip]{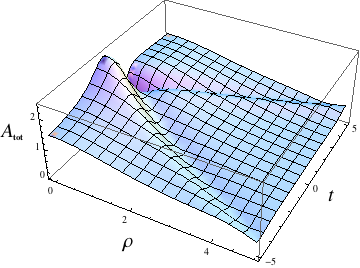}
\\
\includegraphics[width=50mm,clip]{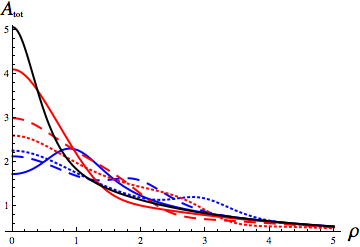}~~
&
\includegraphics[width=50mm,clip]{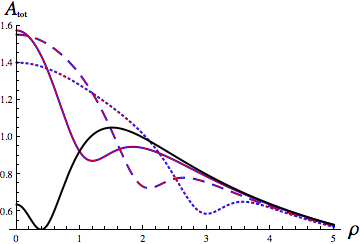}~~
&
\includegraphics[width=50mm,clip]{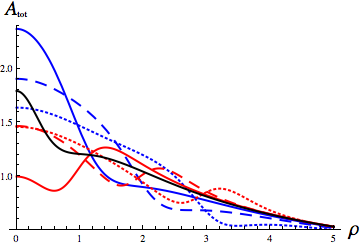}
\end{tabular}
\\
(a) $(k,\theta,q)=(5, n\pi/6,1)$
\\[3mm]
\begin{tabular}{lll}
$n=1$
&
$n=3$
&
$n=4$
\\[-1mm]
\includegraphics[width=5cm,clip]{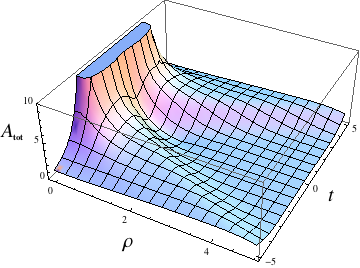}~~
&
\includegraphics[width=5cm,clip]{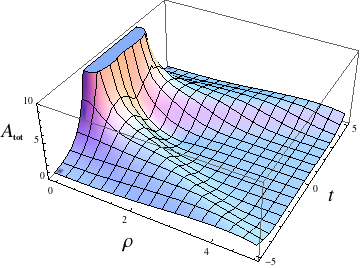}~~
&
\includegraphics[width=5cm,clip]{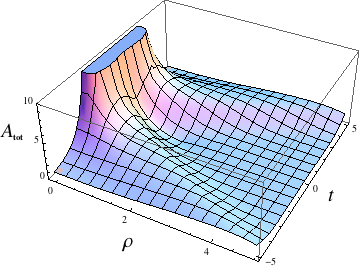}
\\
\includegraphics[width=50mm,clip]{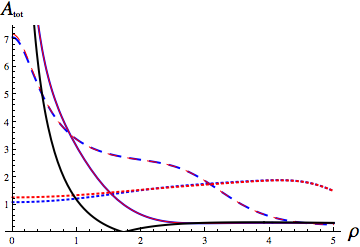}~~
&
\includegraphics[width=50mm,clip]{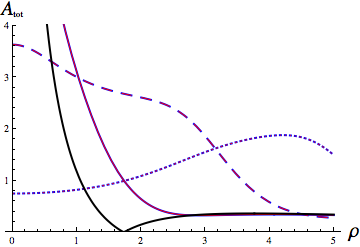}~~
&
\includegraphics[width=50mm,clip]{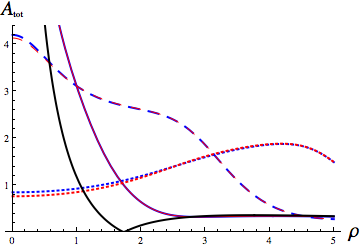}
\end{tabular}
\\
(b) $(k,\theta,q)=(1000, n\pi/6,1)$
 \caption{
Wave propagation in the $t$-$\rho$ plane (upper panels) and 
time evolution of the wave forms (lower panels) 
before and after the reflection at the axis for $d=3$. 
The blue and red curves denote incident waves and reflected waves, respectively. 
The dotted, dashed and solid curves for each color show 
$A_\textrm{tot}$ at $t=\pm3, \pm2, \pm1$ for (a) 
and at $t=\pm5, \pm3,\pm1$ for (b).
The black solid curves show $A_\textrm{tot}$ at $t=0$. 
The blue curves completely coincide with the red curves for $n=3$. 
}
 \label{fig:amplitude-d3}
\bigskip
\end{figure}

\begin{figure}[!t]
\centering
\begin{tabular}{lll}
$n=1$
&
$n=3$
&
$n=4$
\\[-1mm]
\includegraphics[width=5cm,clip]{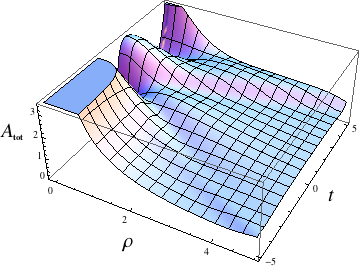}~~
~~&
\includegraphics[width=5cm,clip]{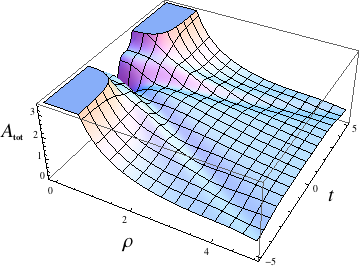}~~
~~&
\includegraphics[width=5cm,clip]{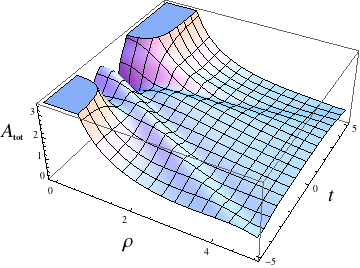}
\\
\includegraphics[width=50mm,clip]{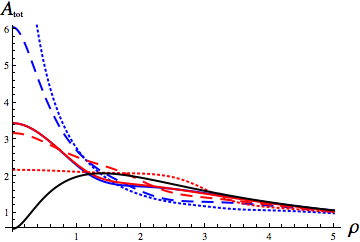}~~
~~&
\includegraphics[width=50mm,clip]{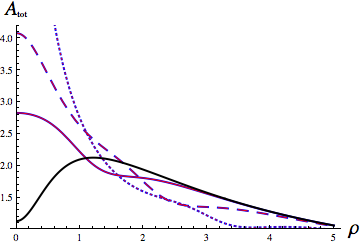}~~
~~&
\includegraphics[width=50mm,clip]{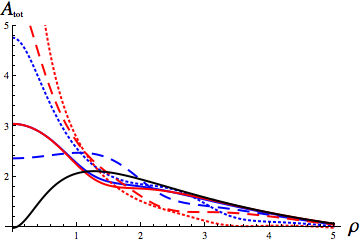}
\end{tabular}
\\
(a) $(k,\theta,q)=(5, n\pi/6,1)$\\[3mm]
\begin{tabular}{lll}
$n=1$
&
$n=3$
&
$n=4$
\\[-1mm]
\includegraphics[width=5cm,clip]{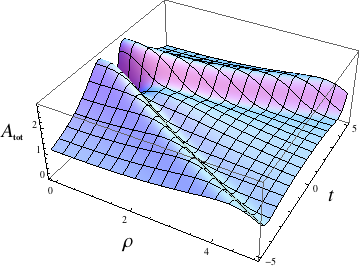}
~~&
\includegraphics[width=5cm,clip]{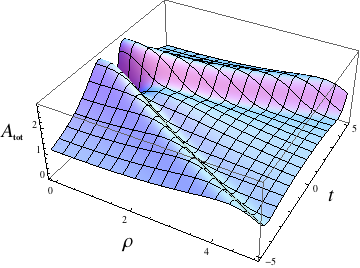}
~~&
\includegraphics[width=5cm,clip]{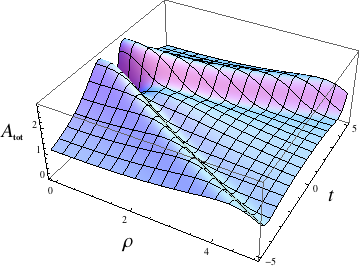}
\\
\includegraphics[width=50mm,clip]{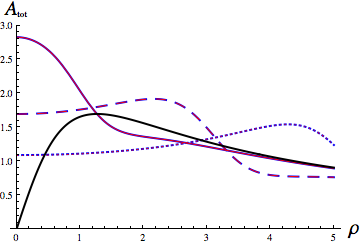}
~~&
\includegraphics[width=50mm,clip]{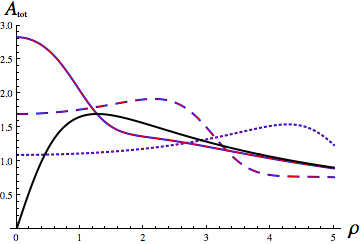}
~~&
\includegraphics[width=50mm,clip]{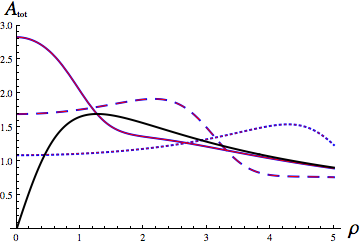}
\end{tabular}
\\
(b) $(k,\theta,q)=(1000, n\pi/6,1)$
 \caption{
Wave propagation in the $t$-$\rho$ plane (upper panels) and 
time evolution of the wave forms (lower panels) 
before and after the reflection at the axis for $d=5$. 
The blue and red curves denote incident waves and reflected waves, respectively. 
The dotted, dashed and solid curves for each color show 
$A_\textrm{tot}$ at $t=\pm3, \pm2, \pm1$ for (a) 
and at $t=\pm5, \pm3,\pm1$ for (b).
The black solid curves show $A_\textrm{tot}$ at $t=0$. 
The blue curves completely coincide with the red curves for $n=3$. 
}
 \label{fig:amplitude-d5}
\bigskip
\end{figure}

Let us see the qualitative behavior of 
the gravitational waves described by the obtained two-soliton solution 
near the axis $\rho=0$. 
Figures~\ref{fig:amplitude-d3} and \ref{fig:amplitude-d5}
display the typical wave forms and their snapshots 
of the total amplitudes, $A_\textrm{tot}$, near the coordinate origin 
for $d=3$ and $d=5$, respectively, 
where we specify $c$ by two real parameters $k$ and $\theta$ as
\begin{align}
c=k\,e^{i\,\theta},
\end{align}
and moreover we set $\theta=n\pi/6$. 
In each figure, we consider the three cases $n=1, 3, 4$ only 
to see the typical picture of reflection. 
Both figures explicitly show that 
the obtained two-soliton solution describes 
the reflectional phenomenon of cylindrical gravitational waves. 
For $k\simeq q$, 
the solution provides the picture of soliton reflection near $t=\rho=0$ 
as seen in Figs.~\ref{fig:amplitude-d3}(a) and \ref{fig:amplitude-d5}(a). 
Note that for $d=5$, 
$A_\textrm{tot}$ becomes larger and larger near the axis 
as $|\,t\,|$ increases in Fig.~\ref{fig:amplitude-d5}(a). 
For $k\gg q$, 
the dependence of $A_\textrm{tot}$ on $\theta$ is quite small 
as seen in Figs.~\ref{fig:amplitude-d3}(b) and \ref{fig:amplitude-d5}(b). 
From these behaviors of the amplitudes, 
we may consider that the two-soliton solutions 
show the propagation of gravitational wave packets which first come into 
the symmetric axis from past null infinity $u=-\infty$, 
and leave the axis after reflection for future null infinity $v=\infty$.

\subsection{Time shift}

\begin{figure}[!t]
\begin{tabular}{lll}
$n=1$
&
$n=3$
&
$n=4$
\\[-1mm]
\includegraphics[width=50mm]{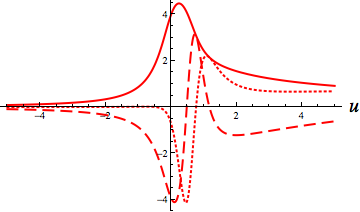}~~
&
\includegraphics[width=50mm]{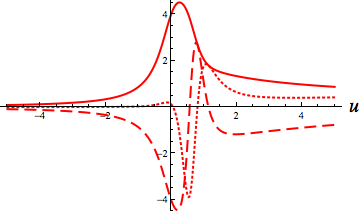}~~
&
\includegraphics[width=50mm]{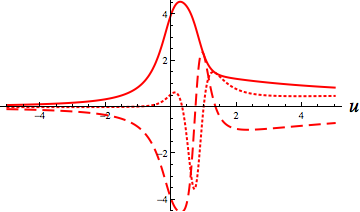}~~
\\
\includegraphics[width=50mm]{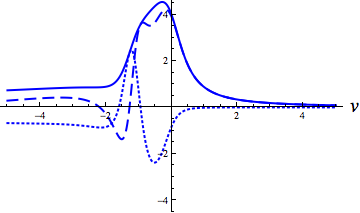}~~
&
\includegraphics[width=50mm]{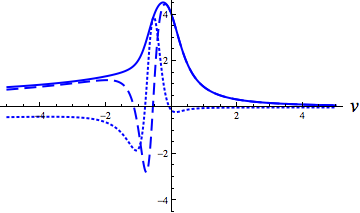}~~
&
\includegraphics[width=50mm]{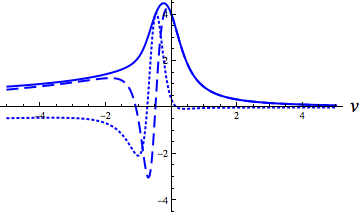}~~
\end{tabular}
\caption{
The ingoing wave amplitudes at $u=-\infty$ (blue curves) and 
the outgoing wave amplitudes at $v=\infty$ (red curves) for $d=3, 5$ 
and $(k,\theta,q)=(1000, n\pi/6,1)$. 
The solid, dashed, and dotted blue-colored (red-colored) curves are 
$A\sqrt{-u}$ $\left(B\sqrt{v}\right)$, $A_+\sqrt{-u}$ $\left(B_+\sqrt{v}\right)$, and $A_\times\sqrt{-u}$ $\left(B_\times\sqrt{v}\right)$, respectively. 
}
\label{fig:timeshift}
 \bigskip
\end{figure}

A time shift phenomenon is known as the nonlinear effect of solitons, 
which means that a wave packet propagates at slower speed than the light velocity 
by its self-interaction when a cylindrical wave collapses near the axis. 
Let us note that the amplitudes for ingoing waves near past null infinity 
$A$ [Eqs.~\eqref{eq:Apast-d3} and \eqref{eq:Apast-d5}] 
and the amplitudes for outgoing waves near future null infinity 
$B$ [Eqs.~\eqref{eq:Bfuture-d3} and \eqref{eq:Bfuture-d5}] 
have the same forms as for $d=1$. 
Therefore, the behaviors near $u=-\infty$ and $v=\infty$ for $d=3,5$ 
is exactly the same as 
for $d=1$ discussed in~\cite{Tomizawa:2015zva}. 
In Fig.~\ref{fig:timeshift}, the blue-colored graphs 
and the red-colored graphs denote 
$\lim_{u\to-\infty}A\sqrt{-u}$ 
and 
$\lim_{v\to\infty}B\sqrt{v}$ 
for $(k,\theta)=(1000,\, n\pi/6)$, respectively, 
where we take $n=1,3,4$ to see the typical behaviors of waves. 
The dashed and dotted blue-colored (red-colored) graphs denote 
$A_+ \sqrt{-u}$ $\left(B_+\sqrt{v}\right)$ and 
$A_\times\sqrt{-u}$ $\left(B_\times\sqrt{v}\right)$ 
at null infinity $u=-\infty$ $\left(v=\infty\right)$, respectively. 
It is worth noting that for large value of $k$, 
both amplitudes near past and future null infinity 
are composed of $\times$ mode waves.

\medskip
To see that a time shift phenomenon happens, 
as was already explained in~\cite{Tomizawa:2015zva}, 
let us consider the incoming massless test particle 
which starts from past null infinity, 
propagates along the null geodesic $v=0$, 
is reflected on the axis $\rho=0$ 
and then propagates to future null infinity along the null geodesic $u=0$. 
An incident wave packet has a peak near $v<0$, 
while a reflectional wave packet has a peak at $u>0$. 
This means that an observer at past null infinity sees 
an ingoing wave packet earlier than an incoming radial photon, 
but at future null infinity he sees the outgoing wave packet after the outgoing photon. 
We may consider that a gravitational wave packet 
can propagate at slower speed than the light velocity.

\subsection{Coalescence and split of solitons}

\begin{figure}[!t]
\begin{tabular}{lll}
$n=1$
&
$n=3$
&
$n=4$
\\[-1mm]
\includegraphics[width=50mm]{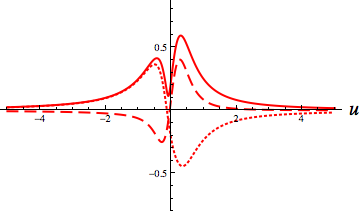}~~
&
\includegraphics[width=50mm]{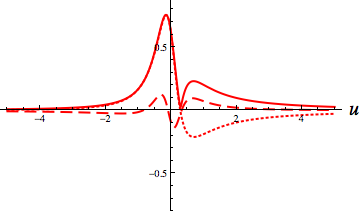}~~
&
\includegraphics[width=50mm]{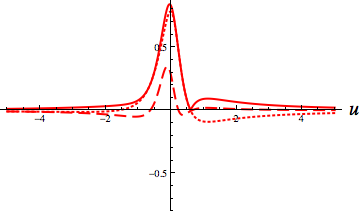}~~
\\
\includegraphics[width=50mm]{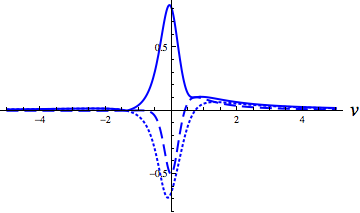}~~
&
\includegraphics[width=50mm]{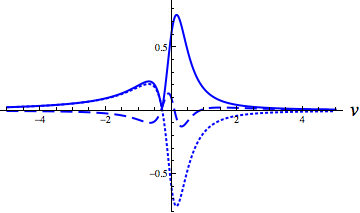}~~
&
\includegraphics[width=50mm]{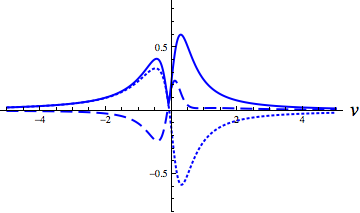}~~
\end{tabular}
\\
(a) $(k,\theta,q)=(5,n\pi/6,1)$
\\[5mm]
\begin{tabular}{lll}
$n=1$
&
$n=3$
&
$n=4$
\\[-1mm]
\includegraphics[width=50mm]{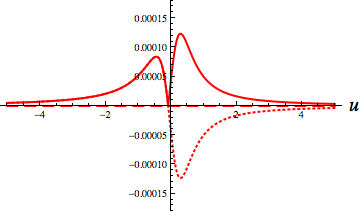}~~
&
\includegraphics[width=50mm]{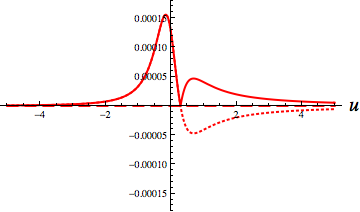}~~
&
\includegraphics[width=50mm]{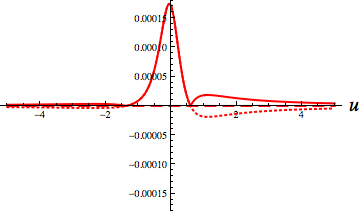}~~
\\
\includegraphics[width=50mm]{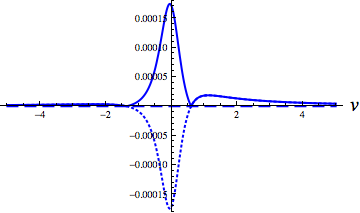}~~
&
\includegraphics[width=50mm]{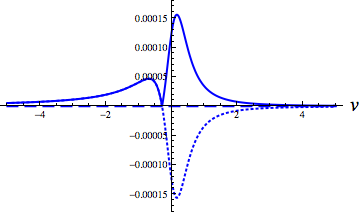}~~
&
\includegraphics[width=50mm]{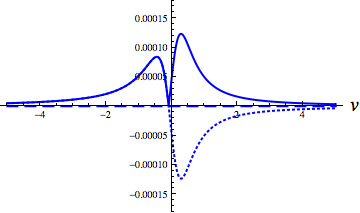}~~
\end{tabular}
\\
(b) $(k,\theta,q)=(1/1000,n\pi/6,1)$
\caption{
The ingoing wave amplitudes at $u=-\infty$ (blue curves) and 
the outgoing wave amplitudes at $v=\infty$ (red curves) for $d=3, 5$. 
The solid, dashed, and dotted blue-colored (red-colored) curves are 
$A\sqrt{-u}$ $\left(B\sqrt{v}\right)$, 
$A_+\sqrt{-u}$ $\left(B_+\sqrt{v}\right)$, and 
$A_\times\sqrt{-u}$ $\left(B_\times\sqrt{v}\right)$, respectively. }
\label{fig:amplitude_nullinfinity}
 \bigskip
\end{figure}

In addition to the time shift phenomena, 
when $k\lesssim q$, physically and mathematically interesting phenomenon 
such as coalescence and split of solitons happens, 
as pointed out in~\cite{Tomizawa:2015zva}. 
As seen in Fig.~\ref{fig:amplitude_nullinfinity}, 
according to the values of $\theta$, 
the ingoing and outgoing waves take various shapes, 
where the blue-colored and red-colored graphs show 
the ingoing amplitude $A\sqrt{-u}$ at past null infinity $u=-\infty$ 
and outgoing amplitudes $B\sqrt{v}$ at future null infinity $v=\infty$, 
respectively, 
and the dashed and dotted graphs show the $+$ mode wave amplitudes 
and the $\times$ mode wave amplitudes, respectively. 
Here, (a) and (b) in Fig.~\ref{fig:amplitude_nullinfinity}
are plotted for the parameters $(k,\theta,q)=(5,n\pi /6,q)$ and $(1/1000,n\pi /6,q)$, 
respectively, where we take $n=1,3,4$. 
From these graphs, we can see that at least, 
either of ingoing and outgoing waves can have two peaks.

\medskip
For $n=3$, 
there are two incident wave packets, 
one with a small peak and another with a large peak near past null infinity, 
and after reflection 
we have two reflectional wave packets, 
one with a small peak and one with a large peak near future null infinity. 
This obviously shows that two gravitational solitons collide with each other, 
which occurs near the axis $\rho\simeq 0$, 
and then the larger one of two solitons overtakes the smaller one 
(see the case $n=3$ in Fig.~\ref{fig:amplitude-d3}).

\medskip
For $n=4$, the incident waves incoming from past null infinity $u=-\infty$ (the blue-colored graphs) 
have two peaks and the reflectional waves outgoing to future null infinity $v=\infty$ (the red-colored graphs) have one peak only. 
As seen from the case of $n=4$ in Fig.~\ref{fig:amplitude-d3}, 
two wave packets seem to change into one wave packet after the reflection at the axis. 
Therefore, this can be interpreted as the coalescence of two solitons.

\medskip
In contrast to $n=4$, for $n=1$, the incident waves (the blue-colored graphs) have 
one peak near past null infinity, 
and the reflectional waves (the red-colored graphs) have 
two peaks near future null infinity. 
From the case $n=1$ in Fig.~\ref{fig:amplitude-d3}, 
one wave packet seems to change into two wave packets 
after the reflection at the axis. 
This phenomenon can be interpreted as the split of soliton waves. 
The phenomena of the coalescence and split do not happen for other solitons 
than ones in general relativity.

\medskip
Finally, we comment on which polarization mode contributes to 
the total ingoing and outgoing amplitudes at null infinity. 
As seen from the behaviors of the dashed and dotted graphs in 
Fig.~\ref{fig:amplitude_nullinfinity}, 
for the small values of $k$ the $\times$ mode only mainly contributes to 
both ingoing and outgoing amplitudes, 
while for the large values of $k$ both the $+$ and $\times$ modes 
contribute to them to the same order.

\section{Conclusions}
\label{sec:4}
In this paper, 
applying the Pomeransky method to a cylindrically symmetric spacetime 
and starting from the Levi-Civit\`a background, 
we have constructed the two-soliton solutions 
with two complex conjugate poles to the vacuum Einstein equations 
with cylindrical symmetry. 
As shown in the previous work, although the Levi-Civit\`a spacetime 
generally includes singularities on its axis of symmetry, 
for the one-soliton solution with $d=3$, 
such singularities can be removed~\cite{Igata:2015oea}. 
In this work, we have analytically shown that 
as for the two-soliton solution with $d=3,5$, 
singularities on an axis entirely disappear in addition to null singularities 
which one solitonic solution with a real pole has in common. 
The regular solutions with $d=3,5$ describe the propagation of 
gravitational wave packets that come into the region near the symmetric axis 
from past null infinity, then leave for future null infinity after reflection at the axis. 
Moreover, we have studied nonlinear effect of solitons 
such as a time shift phenomenon and the gravitational Faraday effect. 
We have seen that these effects are essentially similar to the case of $d=1$, 
which was investigated in~\cite{Tomizawa:2015zva}.

\medskip
Finally, we point out the essential differences of two-soliton solutions with $d=3,5$ from ones with $d=1$.

\medskip
\noindent
(i) Axis of symmetry $\rho=0$: 

For $d=3,5$, 
the \textit{C}-energy density vanishes on the axis $\rho=0$, as for $d=1$, 
although it diverges in the Levi-Civit\`a background spacetime. 
Therefore, since there does not exist any gravitational sources on the axis, 
the two-soliton solution can be physically interpreted as the reflection process 
of gravitational solitonic waves at the axis. 
Furthermore, at late time $t\to \infty$, for $d=3$ the \textit{C}-energy density 
approaches a constant value, while for $d=5$ it becomes an infinitely large value.

\medskip
\noindent
(ii) Timelike infinity $t\to \infty$: 

For $d=1$, the spacetime asymptotically approaches Minkowski, 
and simultaneously both ingoing and outgoing gravitational waves 
fade into the background spacetime. 
The $\times$ mode for the ingoing and outgoing waves becomes dominant at late time. 
On the other hand, for $d=3,5$, the spacetime is not asymptotically Minkowski and both ingoing and outgoing wave amplitudes do not vanish. Moreover, for $d=3\ (c_\textrm{I}\not=0)$ both modes are present, and for $d=5$ the $\times$ mode becomes dominant.

\medskip
\noindent
(iii) Null infinity $v\to \infty$ or $u\to -\infty$: 

Even though the asymptotic forms of the metric at null infinity entirely differ for each of $d$, the asymptotic forms of wave packet are exactly same. 
Therefore, as happens for $d=1$, for $d=3,5$, two gravitational solitons can coalesce
into a single soliton, and also that a single soliton can split into two via the nonlinear effect of gravitational waves. 
Such phenomena cannot be seen for solitons of other integrable equations such as solitons of the KdV equation.

\acknowledgments
This work was partially supported by 
the Grant-in-Aid for Young Scientists (B) (No.~26800120) from Japan Society for the Promotion of Science (S.T.).

\appendix
\section{Definitions}
\label{sec:A}

In this Appendix, 
we provide the well-used definitions on the amplitudes and polarization angles of 
nonlinear cylindrically symmetric gravitational waves, 
which were first used in \cite{Piran, Tomimatsu:1989vw}.

\medskip
The amplitudes of ingoing and outgoing waves with the $+$ mode are defined as, respectively, 
\begin{align}
&A_+=2\,\psi_{,v},
\\
&B_+=2\,\psi_{,u},
\end{align}
and the amplitudes of ingoing and outgoing waves with the $\times$ mode are defined as, respectively, 
\begin{align}
&A_\times=\frac{e^{2\psi}\omega_{,v}}{\rho},
\\
&B_\times=\frac{e^{2\psi}\omega_{,u}}{\rho},
\end{align}
where the advanced ingoing and outgoing null coordinates 
$u$ and $v$ are defined by $u=(t-\rho)/2$ and $v=(t+\rho)/2$, respectively. 
The total amplitudes of ingoing and outgoing waves are defined by
\begin{align}
&A=\sqrt{A_+^2+A_\times^2},
\\
&B=\sqrt{B_+^2+B_\times^2},
\end{align}
respectively, 
and moreover the total amplitude of cylindrical gravitational waves is written as
\begin{align}
A_\textrm{tot}=\sqrt{A^2+B^2}. 
\end{align}
The polarization angles $\theta_A$ and $\theta_B$ for the respective wave amplitudes are defined as
\begin{align}
&\tan 2\theta_A=\frac{A_\times}{A_+},
\\
&
\tan2\theta_B=\frac{B_\times}{B_+}.
\end{align}

\medskip
The vacuum Einstein equations can be written in terms of these quantities as follows:
\begin{align}
&A_{+,u}=\frac{A_+-B_+}{2\,\rho}+A_\times B_\times,
\\
&B_{+,v}=\frac{A_+-B_+}{2\,\rho}+A_\times B_\times,
\\
&A_{\times,u}=\frac{A_\times+B_\times}{2\,\rho}-A_+ B_\times,
\\
&B_{\times,v}=-\frac{A_\times+B_\times}{2\,\rho}-A_\times B_+,
\end{align}
and
\begin{align}
&\gamma_{,\rho}=\frac{\rho}{8}\left(A^2+B^2\right),
\\
&\gamma_{,t}=\frac{\rho}{8}\left(A^2-B^2\right).
\end{align}



\end{document}